\documentclass[12pt]{iopart}
\usepackage{graphicx}

\def\etal{{\em   et\ al.}}
\def\insitu{{\em in\ situ}}

\def\STO{\mbox{SrTiO$_{3}$}}
\def\LAO{\mbox{LaAlO$_{3}$}}
\def\oC{\mbox{$^{\circ}\!$C}}
\def\o{\mbox{$^{\circ}$}}

\begin{document}

\topical[The Structural Analysis Possibilities of RHEED]{The Structural Analysis Possibilities of Reflection High Energy Electron Diffraction}

\author{NJC Ingle, A Yuskauskas, R Wicks, M Paul\footnote{Visiting from Experimentelle Physik 4, Universit\"at W\"urzburg, Am Hubland, 97074 W\"urzburg, Germany}, and S Leung }

\address{Advanced Materials and Process Engineering Laboratory, University of British Columbia, Vancouver, BC, Canada}
\ead{ingle@physics.ubc.ca}

\date{\today}

\begin{abstract}
The epitaxial growth of complex oxide thin films provide three avenues to generate unique properties: the ability to influence the 3-dimensional structure of the film, the presence of a surface, and the generation of an interface.   In all three cases,  a clear understanding of the resulting atomic structure is desirable.  However, determining the full structure of an epitaxial thin film (lattice parameters, space group, atomic positions, surface reconstructions) on a routine basis is a serious challenge.   In this paper we highlight the remarkable information that can be extracted from both the Bragg scattering and inelastic multiple scattering events that occur during Reflection High Energy Electron Diffraction.  We review some methods to extract structural information and show how mature techniques used in other fields can be directly applied to the {\em in-situ} and real-time diffraction images of a growing film.  These collection of techniques give access to both the epitaxially influenced 3 dimensional bulk structure of the film, and any reconstructions that may happen at the surface.

\end{abstract}
\maketitle

\section{Introduction}

The low temperature epitaxial growth of thin films provides a whole new phase space for the structural control of materials compared to bulk solid state chemistry processes.  The ability to affect both kinetic and thermodynamic factors in the growth process with low temperatures and epitaxial stabilization provide a rich play ground to manipulate and control the structure of materials.  There are a wealth of examples in the literature of novel epitaxial structures of metallic, semiconducting, and oxide materials.  Examples include: hcp and bcc Cu and Pd\cite{Wormeester-prl-96}; GaN with cubic\cite{Sun-apl-99}, zinc-blend\cite{Orton-rpp-98}, and wurtzite\cite{Lee-jcg-00} structures; cubic\cite{Salvador-cm-98} and hexagonal\cite{Fujimura-jap-96, Fujimura-apl-96} structures of YMnO$_3$; and layered superlattice materials, such as SrCuO$_2$-BaCuO$_2$\cite{Koster-pc-01}. 

For correlated oxides, where the electronic and magnetic properties are sensitive to small changes in lattice parameter, symmetry, and atom coordination, the ability to grow novel structures could have significant technological implications.   Yonezawa \etal \cite{Yonezawa-ssc-04} have grown V$_{2}$O$_{3}$ epitaxially on two different substrates and succeeded in changing the $a$ lattice parameter by 4\%.  This reduction in lattice parameter caused the stabilization of metallic behavior down to 2K in a system which otherwise would go through a metal-insulator transition at 170K.  In the case of EuO -- a ferromagnetic semiconductor currently being explored as a spin injector material for spintronic applications\cite{schmehl-naturematerials-07} -- the effects of epitaxy\cite{Ingle-prb-08}, in conjunction with doping\cite{Ott-prb-06}, could provide a significant increase to the Curie temperature.  

The perovskite structure -- the basic structure of the well studied high temperature cuprate superconductors, the colossal magnetoresistive manganites, and the ferroelectric titanates -- is one case where evidence of subtle, but critical changes in the structure due to epitaxy have been clearly documented.  For example, SrRuO$_3$, which in its bulk form is orthorhombic due to a small distortion of the Ru-O octahedra compared to the simple cubic perovskite structure,  becomes monoclinic when grown epitaxially on the cubic SrTiO$_3$ substrate.\cite{Gan-jap-99, Vailionis-apl-08}  This subtle change in the structure leads to a uniaxial magnetic anisotropy in the epitaxial films instead of the biaxial magnetic anisotropy found in the bulk.\cite{Gan-jap-99}    

The foundation of epitaxial stabilization is the presence of a specific structure on a surface to act as a template for the growing film.  However, due to the abrupt change in coordination of atoms at a surface, the structure of the surface region can differ from that of the bulk\cite{Henrich-book}.  As such, the nature -- both structural and chemical -- of this surface region, and our ability to understand how it influences an epitaxial interface that is formed on it, is of interest.   

The  external manipulation of the doping levels of a superconductor via an epitaxial interface with a ferroelectric\cite{ahn-nature-03} is one recent example of the potential application of epitaxial interfaces. A second is the non-bulk-like charge states at the interface of polar films of LaAlO$_3$ and LaTiO$_3$ on non-polar SrTiO$_3$ substrates\cite{Ohtomo-nature-02, Ohtomo-nature-04}.  The properties of these polar/non-polar interfaces are reminiscent of the classic semiconductor 2D electron gasses, and basic electronic devices are already being fabricated\cite{Caviglia-nature-08}.  However, a detailed understanding of the generated interface, and the mechanism responsible for the metallic interface is not yet clear.   

In addition, epitaxial growth also provides a means of purposefully generating highly ordered and controlled surfaces.  Theoretical work has suggested that the electronic and magnetic properties of polar surfaces such as MgO(111) on Ag(111), would be very different from their bulk materials properties\cite{Arita-prb-04}.  Due to the varied possibilities of reconstruction on a polar surface, such as defect formation or species absorption, atomic reconstruction, and even faceting, experimental verification of such work requires a detailed view of both the surface chemistry and surface structure.  

In order to better understand the potential of epitaxy and to take full advantage of its capacity to generate novel structures, interfaces, and surfaces with unique properties, the ability to fully explore the structure (unit cell, space group, atom positions) of epitaxial films is a significant part of the puzzle.   Ideally structural analysis would be available \insitu\ and would provide information not only on the bulk structure of the film, but also give access to surface structures.  

Due to the intrinsically highly-ordered nature of epitaxial films, diffraction is well suited to provide this structural information.  The direct results of experimental diffraction measurements are the unit cell parameters, the space group (or a small selection of possibilities) and the intensity data.  Full structural knowledge, which would include the atomic positions, is not immediately available from standard diffraction techniques due to the lack of phase information, although there are many indirect and direct methods that can be applied\cite{Saldin-jpcm-08}.  

However, using the standard structural determination techniques and tools on thin films is a challenging problem.  Primarily this is because of the small quantity of the material of interest (films are often $<2000$\ \AA\ thick) and the weak interaction of x-rays and matter.   This is made worse for oxides, due to the low cross-section for low Z elements.  Furthermore, the two-dimensional nature of the film makes it impossible to reach a significant portion of reciprocal space.  This shortens the list of possible observed and unobserved peaks and their intensities making standard indirect and direct methods for atom position determination a serious challenge.  

Synchrotron based x-ray diffraction can help to overcome the weak interaction issue by increasing the brilliance of the x-ray beam by a factor of approximately $10^{12}$ over lab-based x-ray generation\cite{Elements-of-Modern-X-Ray-Physics-Als-Nielsen}.    This increased brilliance allows for the detailed exploration of surface-related x-ray diffraction effects via crystal truncation rods (CTR)\cite{Robinson-rpp-92}.  CTRs are created in reciprocal space, instead of a 3-dimensional reciprocal lattice, due to the breaking of the crystal periodicity perpendicular to the surface of the film.   The CTRs can provide information about the vertical structure of the surface, and can provide a clear distinction between bulk structure and that of the surface if there is a change in the lateral periodicity of the surface with respect to the bulk. This can occur due to surface reconstructions or surface adatoms, and leads to CTRs that have contributions from both the bulk and surface regions, and superstructure rods which are a direct consequence of just the surface.  Fong and Thompson\cite{Fong-armr-06} highlight recent work using synchrotron radiation for \insitu\ studies of bulk and surface properties of perovskite ferroelectrics thin films. 

The weak interaction issue with x-ray diffraction studies of thin films can also be solved by using electrons as the probe particle.  Transmission electron microscopy-based diffraction methods can provide a wealth of information on bulk and interface structures of thin films (see, for example, \cite{Marshall-jap-99, Li-jmr-98, Ohtomo-nature-02}), and can also provide chemical information\cite{muller-naturematerials-09}.   However, it requires significant sample preparation which is destructive to the sample being studied, and does not allow access to surface related effects.   An \insitu\ technique allowing both the determination of the full crystallographic structure (space group, lattice parameters, and atomic positions) and surface related crystallographic structures would be a powerful and desirable tool; the ability to obtain this type of information in real-time during film growth would be an added bonus.

The standard tools for \insitu\ structural determination are Low Energy Electron Diffraction (LEED) and Reflection High Energy Electron Diffraction (RHEED).  LEED probes the surface layer of atoms to be studied, and as such it is the primary tool to uncover the crystallographic structure (with immediate visualization of the symmetry information) of the surface and any surface reconstructions that are present\cite{Heinz-rpp-95}.  However, the front view geometry of the LEED measurement limits its use as a real-time growth monitor.  

The grazing incidence geometry of RHEED makes it well suited to real-time monitoring during thin film growth, as the source of electrons and detection scheme are physically removed from growth sources and other ancillary growth equipment.  Minor design modifications even allow its use under poor vacuum conditions, of up to 50 Pa\cite{Rijnders-apl-97}. The geometry also provides a high level of surface sensitivity for the 15-30 keV electrons that are normally used.  RHEED equipment is found in the vast majority of epitaxial growth chambers, primarily because the basic RHEED equipment, its implementation in a growth chamber, and the initial interpretation of the patterns are fairly straightforward.  

Due in part to the geometry, and also to the various inelastic scattering effects that come along with electron diffraction, there is a wealth of information that can be collected and extracted from RHEED based diffraction.  This information runs the gamut from topological information all the way to full structural information of both the bulk and surface region.   There are a number of excellent review papers and books that cover the many details of RHEED pattern formation, and the wide range of growth and topological information that can be accessed from RHEED\cite{Mahan-jvs-90,Daweritz-sst-94, Mitura-srl-99, braun-book, wang-book, ichimiya-book, larson-book}.  

This paper concentrates on the possible approaches with RHEED to obtain detailed structural information.  The need for this type of crystallographic information is strongly driven by the nature of complex oxide materials -- complex in structure and composition --  where expitaxially driven changes in atomic coordination and small changes in bond lengths can have dramatic effects on their properties.  In particular, we highlight recent techniques that allow rapid data collection so that they can be used during growth processes.  Furthermore, we are also primarily interested in techniques that can either be interpreted directly from the image, or with fairly straight forward data manipulation and simulation tools.

\section{Standard RHEED information}

\begin{figure}
\begin{center}
\includegraphics[width=0.77\textwidth]{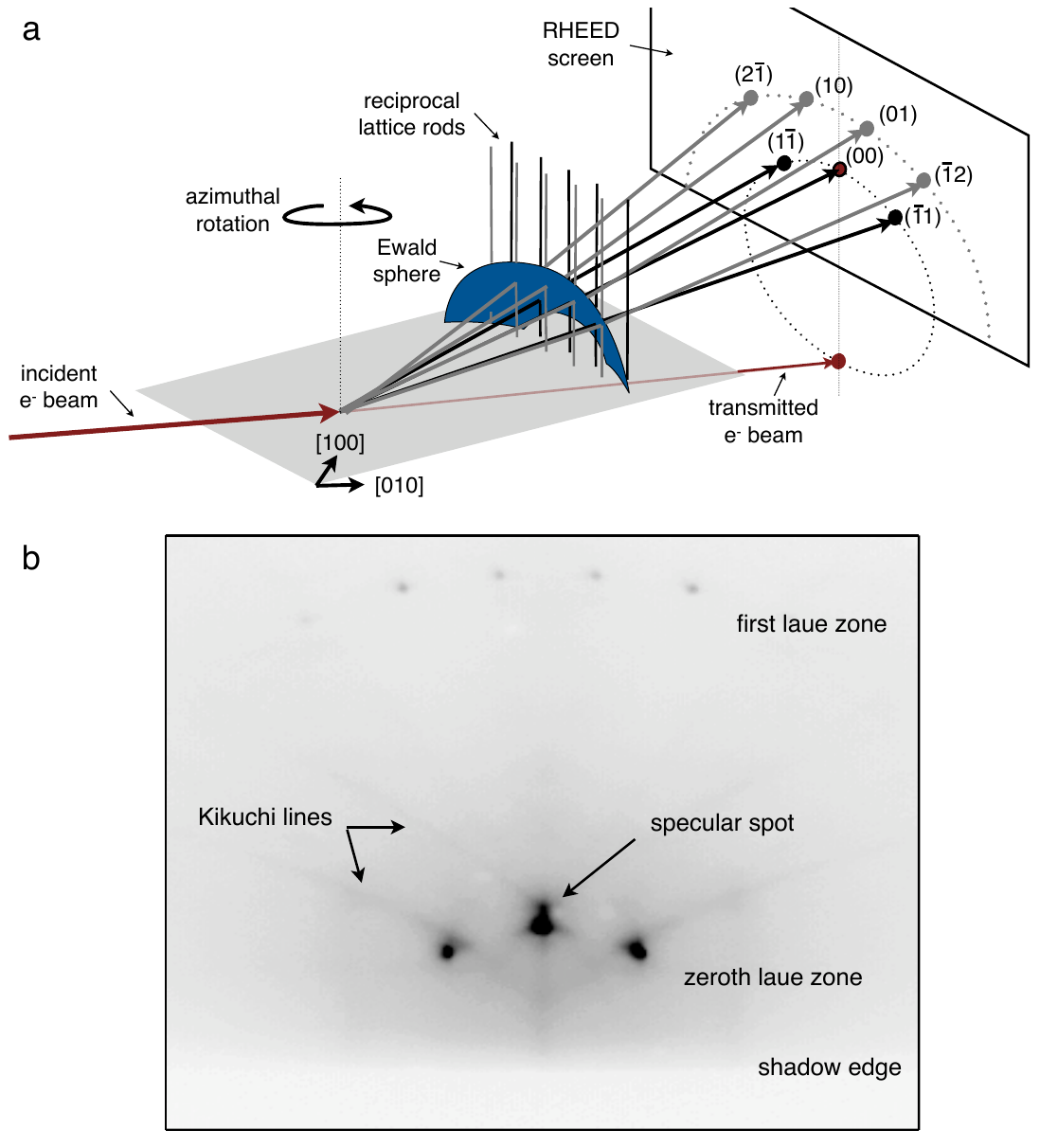}
\caption{a) The basic geometrical setup of RHEED with the Ewald sphere construction highlighted. b) An exemplary RHEED image with the shadow edge, specular spot, Laue zones, and Kikuchi lines identified.\label{basic-RHEED}}
\end{center}
\end{figure}

The geometry of RHEED is shown in  \fref{basic-RHEED}a, along with a typical RHEED image in \fref{basic-RHEED}b.  For an ideal flat surface, the RHEED diffraction image is a gnonomic projection of the intersection of the crystal truncation rods (CTRs) and the Ewald sphere, and as such is a series of high intensity points on a  dark background.  These diffraction maxima are from single scattering Bragg diffraction events.  Mahan\cite{Mahan-jvs-90} provides a review of the basic geometrical arguments for RHEED images.  On most surfaces, however, defects such as step edges or islands generally limit the distance over which long range order, and hence coherent interference, are maintained.  Broadening of the CTRs from these effects leads to streaky RHEED patterns.  The form of these streaks, and their dependence on azimuthal angle can help distinguish various forms of island growth, and rotationally disordered surfaces (see, for example, Ref. \cite{ichimiya-book}).   

\begin{figure}
\begin{center}
\includegraphics[width=0.77\textwidth]{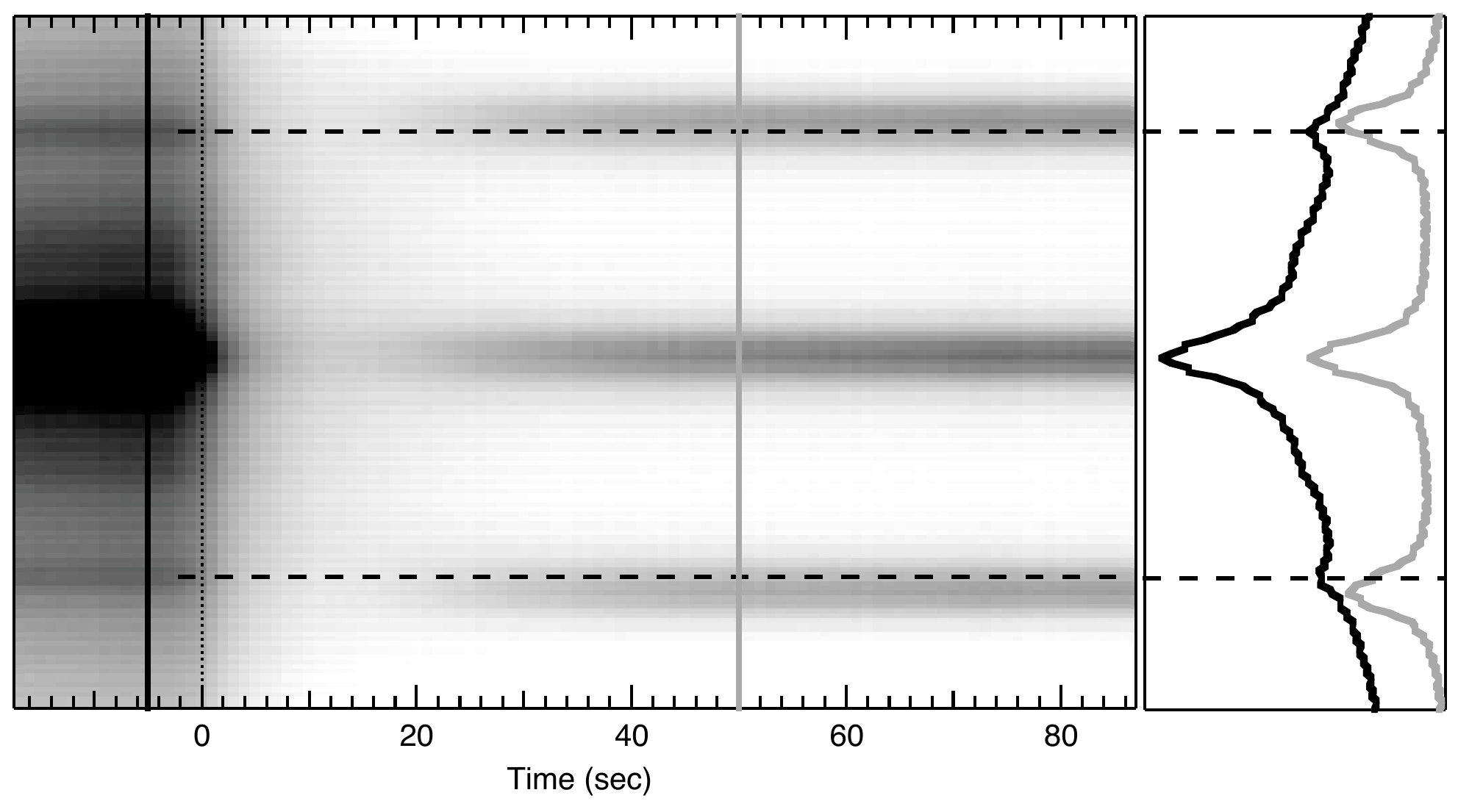}
\caption{Integrated RHEED-image profile as a function of time of EuO grown on LaAlO$_3$, along the [110] direction of the pseudo-cubic unit cell of LaAlO$_3$.  At $t<0$, the profile is that of the LaAlO$_3$ substrate.  At $0<t<20$ seconds, which corresponds to approximately 5\AA\ of growth, the RHEED image completely disappears, except for some minor intensity in the spectrally reflected spot.  At $t>20$ seconds a strong RHEED pattern reemerges and shows the bulk d-spacing for (100) EuO, which is smaller than that of the (110) pseudo-cubic d-spacing of LaAlO$_3$. The right hand side of the figure shows profiles from $t=-5$ sec and $t=50$ sec.\label{RHEED-timeseries}}
\end{center}
\end{figure}

The temporal changes in the RHEED images are a rich source of information about the growth process.  When the diffraction patterns along a high symmetry crystal directions are collected during the growth of the film, they can be arranged to highlight growth mode changes as a function of film thickness.   \Fref{RHEED-timeseries} shows a series of RHEED images where the area of the images that contain the first Laue zone have each been integrated perpendicular to the shadow edge to generate a line profile, and then plotted as function of time.  This particular time series is of EuO grown on LaAlO$_3$ with an epitaxial relationship of EuO(100) on LaAlO$_3$(012) with EuO[100]//LaAlO$_3$[$\bar{1}\bar{2}1$].  The first 18 seconds in \fref{RHEED-timeseries} is prior to the start of growth and shows the bright specular spot ($00$), and the ($2\bar{1}$) and ($\bar{1}2$) diffraction spots of the the first Laue zone along the [$\bar{1}\bar{2}1$] direction of LaAlO$_3$.  At the zero in time the shutter is opened and growth is commenced.  This style of RHEED data presentation shows immediately that a disordered interface layer grows directly on the substrate, and that as the film thickness increases, a single crystal-like growth commences.  At times greater than about 20 seconds, the specular spot (00), and the (10) and ($\bar{1}0$) diffraction spots of the EuO gain intensity.  Additionally it shows that there is a change in d-spacing perpendicular to the direction of the incoming beam of the single crystal-like film compared to the substrate. In this case, the bulk d-spacing for EuO of 5.14\AA\ is obtained in the film immediately on top of the disordered layer, as compared to the 5.35\AA\ d-spacing of the LaAlO$_3$ substrate.

RHEED has also been used extensively to track the thickness of materials grown in a layer-by-layer fashion.\cite{braun-book}  In the simple picture, the order and then disorder of a completed and then partially completed growth layer generates an oscillatory behavior in the intensity of the specularly reflected beam.  These oscillations are seen most clearly at low angles of incidence, and often several degrees away from a high symmetry direction to minimize the influence of multi-scattering events.\cite{Shin-jvs-07}  One of the key applications of these thickness oscillations is the calibration of growth rates.  They have been used to carefully study the absorption-controlled growth mechanism of oxides, such as EuO\cite{steeneken-thesis}.   The oscillations also allow a very precise method to control the growth of superlattice structures\cite{Koster-pc-01}, and to study the diffusion related relaxation of the growth process \cite{Blank-jcc-00}.  The growth of the recent \STO/\LAO\ superlattices, with control of layers down to half a unit cell is the most recent example of the striking control that is afforded by the realtime monitoring of such oscillations \cite{Ohtomo-nature-04,Thiel-science-06,Huijben-np-06}.

\begin{figure}

\begin{center}
\includegraphics[width=0.77\textwidth]{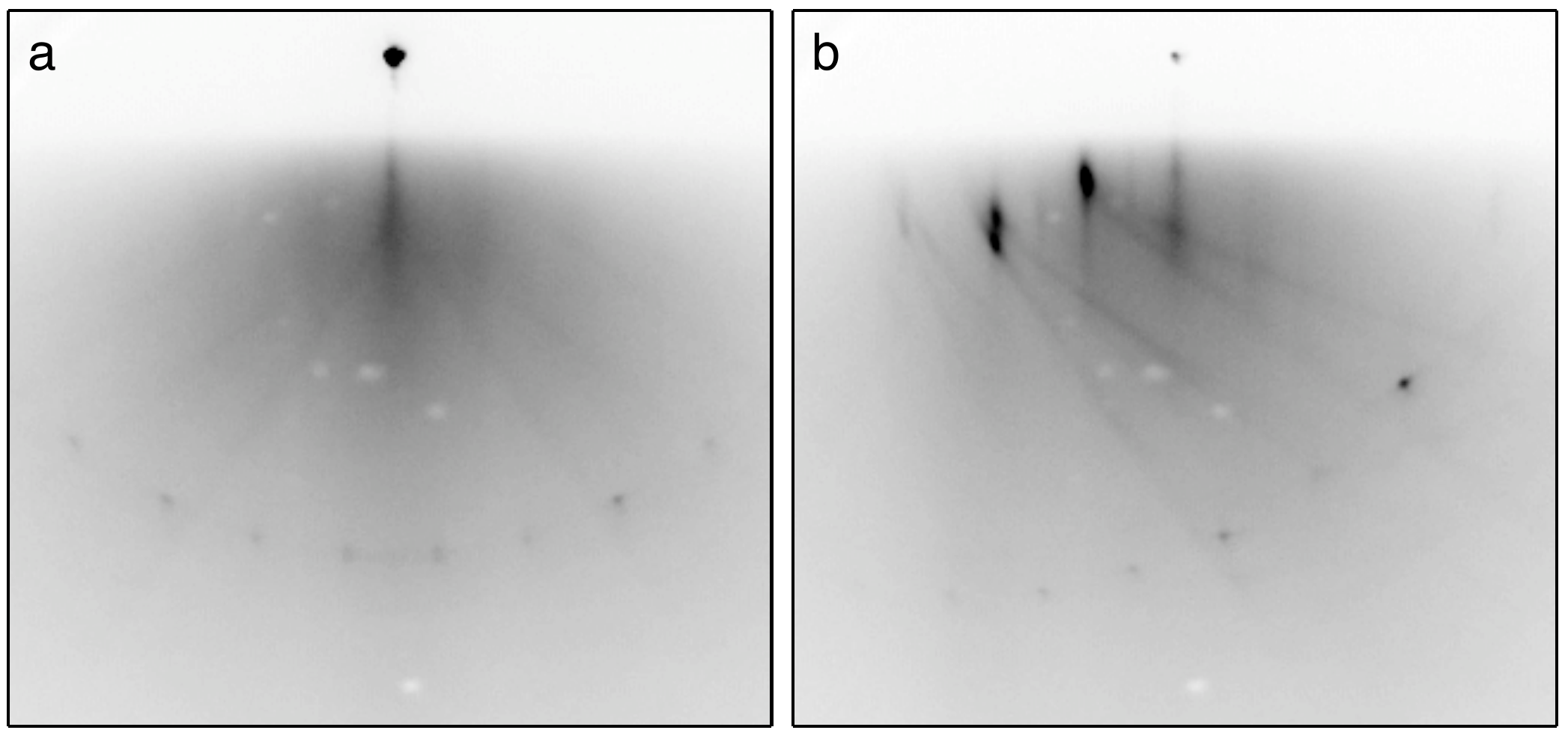}
\caption{RHEED images of SrTiO$_3$.  a) Along the (001) zone axis, the 0th zone Bragg diffraction is barely visible over the large background intensity.  b) 4\o off the (001) zone axis and there is significant enhancement where the Kikuchi pattern and the 0th zone single scattering Bragg diffraction overlap. \label{STO_RHEED_kikuchi}}
\end{center}
\end{figure}

In addition to the Bragg diffraction that is the primary information in RHEED images, there can also be strong patterns generated by the subsequent diffraction of inelastically scattered electrons.  These patterns, called Kikuchi patterns and discussed in detail later, can be quite sharp and contain significant intensity in the case of crystals with very high bulk and surface order.   The patterns are easily distinguishable from the single scattering Bragg diffraction intensity because they move in a continuous and coincident manner when the crystal is rotated. As such, they can be used to align the azimuthal direction of the sample with respect to the incoming electron beam to a high accuracy.  These Kikuchi patterns are also traditionally used to aid sample alignment in Transmission Electron Microscopy.  One major complication associated with the Kikuchi patterns is that there can be significantly enhanced intensity where the Kikuchi patterns and the Bragg diffraction patterns overlap.  This effect is called a surface-wave resonance \cite{wang-book}, and can lead to an order of magnitude increase in the intensity.    In fact, the RHEED patterns from some samples, such as high quality SrTiO$_3$ (\fref{STO_RHEED_kikuchi}), are so dominated by these enhancements that it can be quite hard to clearly see the Bragg diffraction pattern.  

\section {RHEED as a surface symmetry tool}

\begin{figure}

\begin{center}
\includegraphics[width=0.47\textwidth]{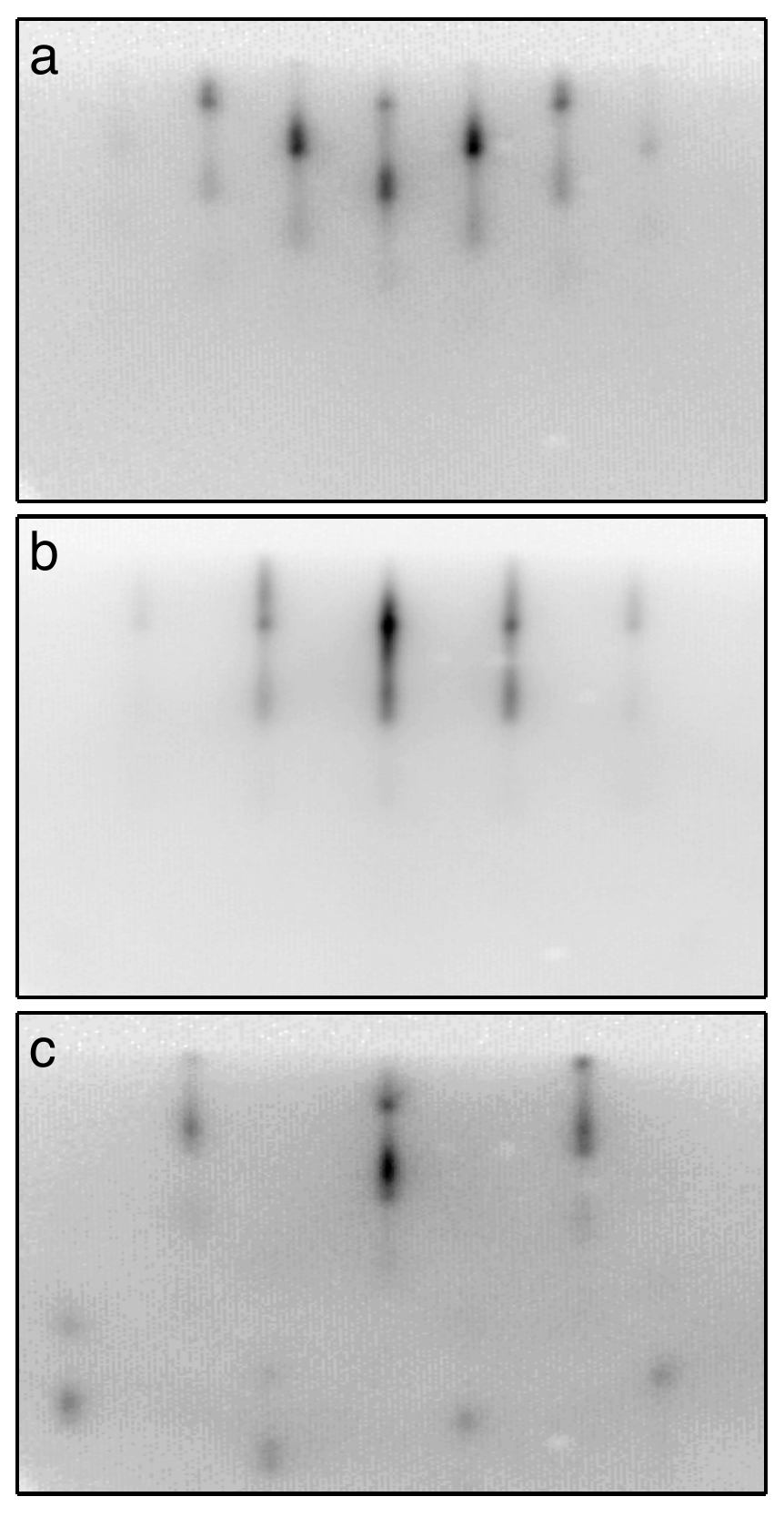}
\caption{RHEED images of a quasi-3D EuO film on LaAlO$_3$ at (a) 0\o\ to the LaAlO$_3$ pseudo-cubic (100) direction, (b) 45\o\ , and (c)  26.56\o.  Diffraction from the first Laue zone, with expected lateral spacing, can be seen in (c). \label{EuO-azimuthal-RHEED}}
\end{center}
\end{figure}

Beyond information on surface morphology and time-dependent growth information, RHEED also allows the underlying in-plane symmetry of the CTRs to be extracted.  Collecting RHEED images as the azimuth angle of the sample is varied allows the high symmetry crystal directions to be found.  \Fref{EuO-azimuthal-RHEED} shows the RHEED images of a quasi-3D 1000\AA\ EuO film grown on LaAlO$_3$(120) at 0\o, 45\o, and 26.56\o\ with respect to the LaAlO$_3$ [$\bar{1}\bar{2}1$] direction.  These images allow d-spacing to be determined for the (100), (110), and (120) planes respectively, and hence the in-plane unit cell lattice parameters of the EuO film.  Use of the known lattice parameters of the substrate prior to film growth allows for the accurate calibration of the RHEED geometry and detection system.

\begin{figure}

\begin{center}
\includegraphics[width=0.57\textwidth]{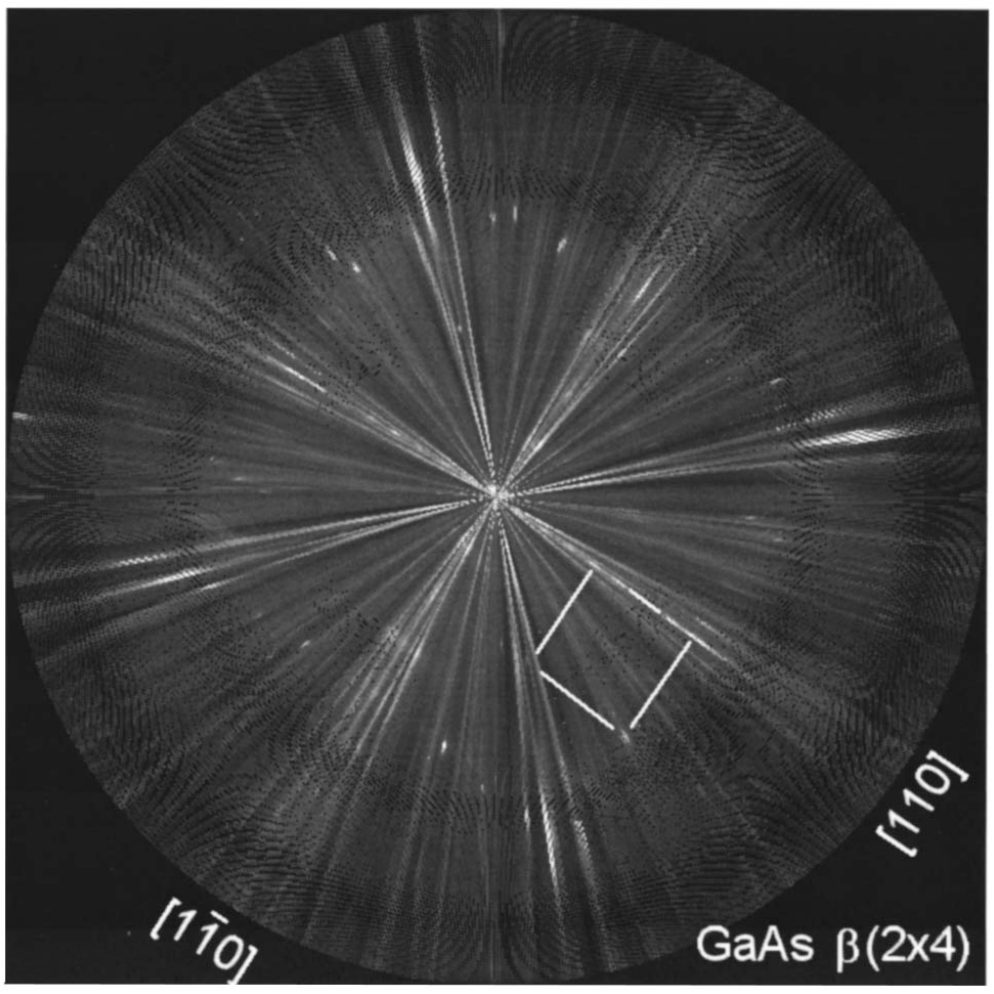}
\caption{Azimuthal scan of a GaAs $\beta(2\times4)$ reconstructed surface. The main azimuths and the surface unit cell are indicated. Reprinted from \cite{Braun-jvs-98}\label{braun} }
\end{center}
\end{figure}

The azimuthal angular spread of the high symmetry zone axes allows the basic in-plane symmetry to be determined.  Braun \etal\cite{Braun-jvs-98} demonstrated one method of more directly viewing this symmetry information.  By arranging, as a function of azimuth angle, a single horizontal slice through the specular spot of each RHEED image, a map of intensity versus the azimuthal angle is obtained.   This map ideally produces an image similar to that of LEED, and as such gives a direct visualization of the symmetry shown by the CTRs.   \Fref{braun} shows the a GaAs $\beta(2\times4)$ reconstructed surface from Braun\etal\cite{Braun-jvs-98}.

When interpreting RHEED with the Ewald sphere construction method, the RHEED image is a gnomonic projection of the Ewald sphere onto the flat phosphor screen. All high intensity points in the RHEED image indicate locations where the Ewald sphere and the CRT's intersect.   As such, a simple transformation, relative to the shadow edge of the RHEED image,  can be applied to each point in the RHEED image to map the diffraction intensity onto the surface of the Ewald sphere in 3-dimensional space.  By collecting RHEED images as a function of azimuthal angle and transforming each image back to the Ewald sphere which is now rotating with respect to the sample's reciprocal lattice, a dense intensity map is built up which will reproduce the CTRs in 3-dimensions.  In scattering language, each point in the RHEED image is a different scattering vector ($\vec{q} = \vec{k_f} - \vec{k_i}$) with respect to the reciprocal lattice of the sample.  As the sample is rotated, a large range of scattering vectors are swept out in all 3 dimensions.  According to the Von Laue condition, intensity will be seen at those scattering vectors which match the CTRs.  

When integrating this 3-dimensional intensity map over the direction perpendicular to the samples surface, an image can be obtained that displays the crystallographic symmetry of the CTRs.  This is a much higher signal-to-noise version of the visualization presented by Braun \etal\cite{Braun-jvs-98}.  Some limited structure factor information, such as systematic absences, can also be seen.  Experimentally, a precession-free rotation of the sample provides a stable shadow edge and specular spot, which readily allows this transformation to be done in real-time.  

\begin{figure}

\begin{center}
\includegraphics[width=0.97\textwidth]{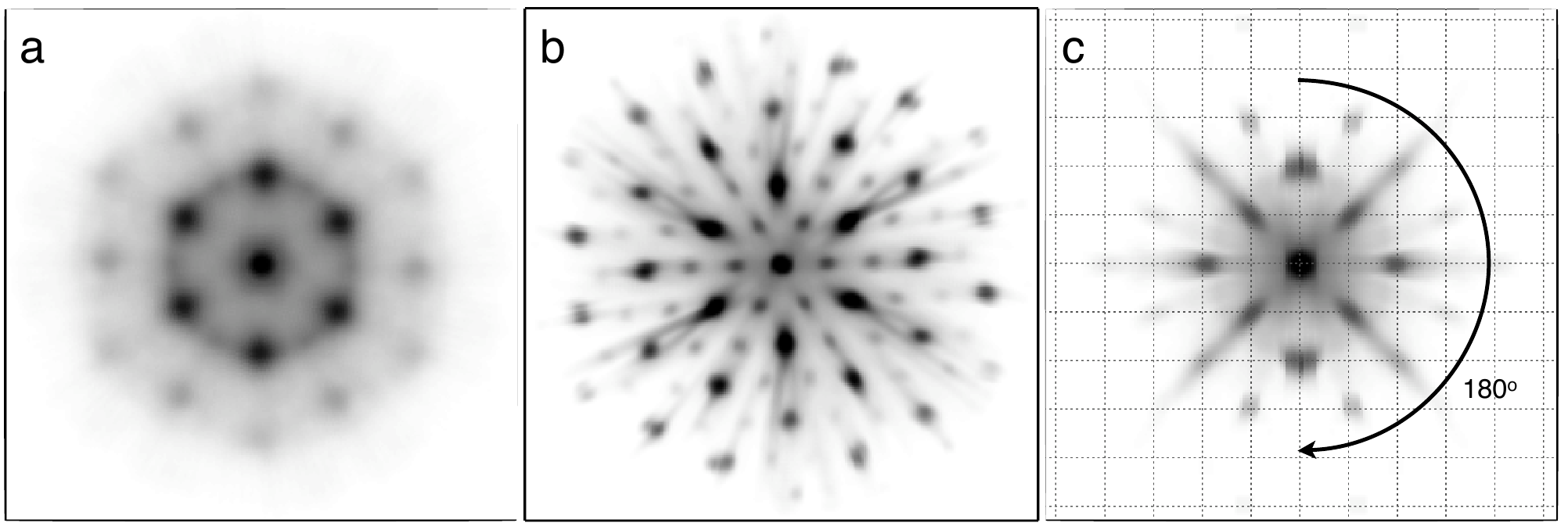}
\caption{A reciprocal lattice visualization from RHEED data of (a) Au deposited on a mica substrate, (b) an O-terminated ZnO ($000\overline{1}$) substrate showing a $R30^{\circ}(\sqrt{3}\times\sqrt{3})$ reconstruction, and (c) a SrTiO$_3$ (001) substrate with the reciprocal lattice spacing indicated by the grid. \label{Au_ZnO_STO}}
\end{center}
\end{figure}

\Fref{Au_ZnO_STO}a shows the reciprocal lattice visualization of RHEED data collected over an azimuthal angle range of 360$^{\circ}$ of a Au epitaxial film grown on a mica substrate.  Au has a cubic unit cell ($Fm\overline{3}m$), so the 3-fold symmetry seen in the reciprocal lattice visualization immediately shows that the film grew with the close packed (111) plane parallel to the mica surface.   To further highlight the information that can be quickly extracted from these RHEED generated surface reciprocal lattice images, \fref{Au_ZnO_STO}b shows the 360$^{\circ}$ data from a ($000\overline{1}$) O-terminated surface of ZnO ($P6_3mc$), taken directly after heating to 700\ $^\circ$C in vacuum.  In this image superstructure rods are clearly seen and indicate a $R30^{\circ}(\sqrt{3}\times\sqrt{3})$ reconstruction is present on this surface.\cite{markus-unpublished}  The radially directed streaks of intensity which start at the \{$10l$\} CTRs are due to the presence of Kikuchi lines in the RHEED images.  

We have found that this visualization method only works well if the intensity of the Bragg diffraction is significantly larger than that of the Kikuchi patterns.  If it is not, the intensity from the   Bragg diffraction will not be the prominent feature of the images.   \Fref{Au_ZnO_STO}c shows the transformed RHEED data of a 180$^{\circ}$ scan of a SrTiO$_3$ (001) substrate.  Generally, SrTiO$_3$ substrates show very strong Kikuchi patterns, which significantly degrade the transformed image.  This particular visualization was built from RHEED imaged collected on poor quality substrates which minimized the influence of the Kikuchi patterns.  SrTiO$_3$ has a cubic ($Pm\overline{3}m$) perovskite structure with a reciprocal lattice spacing of 0.256\ \AA$^{-1}$, as indicated by the grid on \fref{Au_ZnO_STO}c.  As expected, the \{$01l$\} CTRs are missing from the image because their structure factor is very small.   

The spectroscopic make-up of the electrons reaching the RHEED screen has been explored in some detail by a number of researchers\cite{Horio-ass-96, Staib-jcg-99, Nakahara-ass-03}.  They have shown that the elastic peak is generally well separated from the structured loss distribution, which is dominated by surface plasmons, especially at low incident angles\cite{Nakahara-ass-03}.  For the cases of Si (111) \cite{Nakahara-ass-03}, SrTiO$_3$ (001), GaAs (100), and InP (100)\cite{Staib-jcg-99} filtering out electrons that lost more than ~5eV minimizes  the influence of the inelastically scattered electrons from the RHEED image.  This energy filtering removes the diffuse background, Kikuchi lines, and surface resonance features allowing for a more accurate structure determination from the Bragg peaks.  Furthermore, superstructure peaks due to possible long range ordering of the surface are better detectable and higher quality visualizations of the CTRs are generated.  

\section {RHEED as a surface structure tool}

RHEED has been a very attractive tool to explore and understand the large range of reconstructions present on the surfaces of semiconductors, especially under various growth conditions.  As discussed in the previous section, the unit cell of the surface reconstruction is a fairly straightforward piece of information to obtain from RHEED.  However, in order to determine the atomic positions of atoms on the surface, information must be extracted from the intensity of the RHEED image. 

Conventionally,  integrated intensities of various Bragg diffraction spots  -- often chosen so that inelastic scattering effects do not dominate -- are collected as the incident beam angle is varied in the range of $1$\o\ to $\sim5$\o\ for a specific azimuthal angle (see, for example \cite{vonGlan-ss-97, Xie-jcp-03, Hashizume-prl-94, ohtake-ass-03}).  Averages of symmetrically equivalent spots are often also included in the data.  An alternative approach is to vary the azimuthal angle at a specific incident beam angle and integrate the intensity from the specular spot \cite{Mitura-prb-96}.   These intensity versus angle plots are then compared with dynamical calculations of the intensity of high energy electrons reflected from the surface.   The dynamical calculations are multi-beam calculations for elastic scattering and are formulated for the case of two-dimensional periodicity parallel to a surface\cite{maksym-81, Ichimiya-83, zhao-88, ichimiya-book}.  Calculations of all possible atomic configurations consistent with the specific reconstructed surface unit cell need to be compared to the experimental data to determine which atomic configuration provides the best fit.  

A slightly simplified version of this multi-beam approach is also used, called the one-beam method.\cite{Ichimiya-srl-97, ichimiya-book}  This method is based on the idea that if an azimuthal angle is chosen away from a crystallographic symmetry axis, the incident beam will see each atomic layer as an approximately homogeneous continuum.  The variation of the incident beam angle will then primarily probe the changes in the crystal potential in the direction normal to the surface.  Therefore, the specular spot RHEED intensity data can be compared to dynamical calculations of models that  include only the interlayer distances normal to the surface and the atomic densities of those layers as parameters.

These types of trial-and-error procedures can be very demanding when the model parameter space is large, or there is no good initial guess for the surface structure.  However,  recent papers using LEED for surface reconstruction studies have extended the possible approaches beyond the conventional dynamical analysis methods by the use of the Patterson Function, an application from the kinematical methods of x-ray diffraction.\cite{Chang-prl-99, Wu-prl-01, Rogero-prb-03, Wang-prb-05}   The method takes the sum of transforms of LEED I-V spectra over multiple incident directions and momenta transfers which minimizes the effects of multiple scattering and can produce an artifact-free Patterson function of the surface region.

A Patterson function is a real space function calculated by taking the the Fourier transform of measured diffraction intensities in reciprocal space,
\[ P(\vec{r}) = \sum_{hkl} |F_{hkl}|^2 \exp(-i\vec{q}\cdot \vec{r}). \]
It is equivalent to an autocorrelation function of the electron density of a unit cell and so provides information about the interatomic vectors between atoms in a unit cell.  Although the Patterson function approach is not a direct structural determination tool, it allows significant insight into the location and arrangement of atoms within the unit cell.\cite{Robinson-rpp-92} Following the ideas behind the use of the Patterson function on LEED data -- large data sets and multiple incident angles --  it has recently been applied to RHEED data.  

\begin{figure}
\begin{center}
\includegraphics[width=0.57\textwidth]{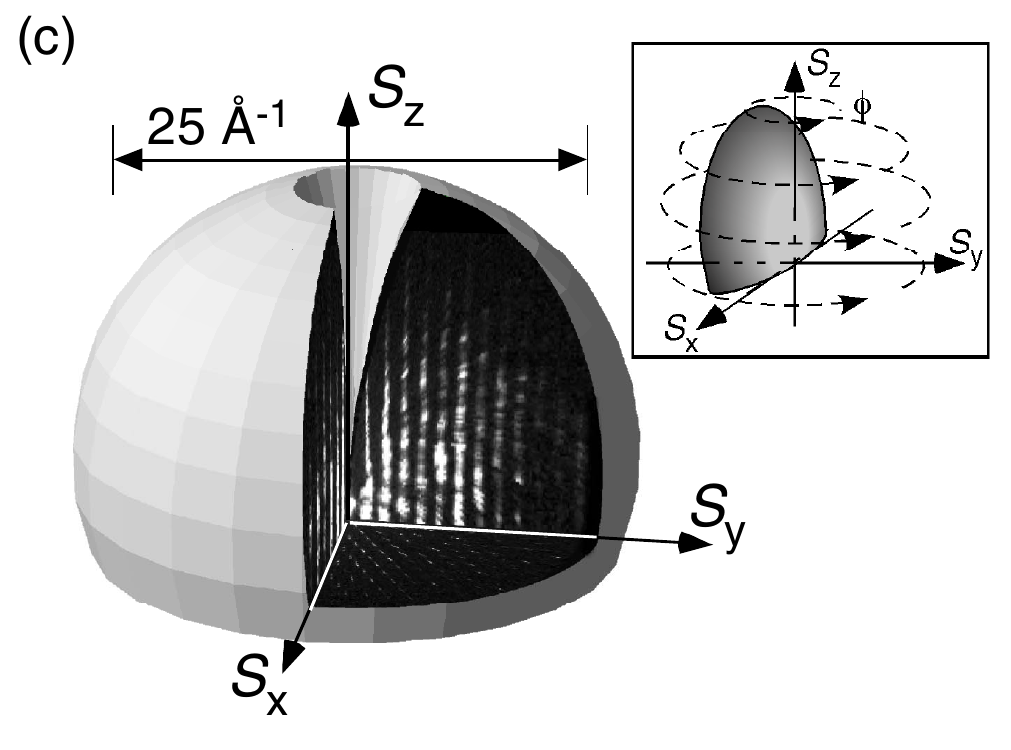}
\caption{Observable volume in the reciprocal space, where a quadrant has been cut to show internal sections, and $\vec{s}$, ($s_x$, $s_y$, $s_z$) is the scattering vector. (Inset) Azimuthal rotation of the Ewald sphere.  Reprinted from \cite{Abukawa-prl-06} \label{abukawa}}
\end{center}

\end{figure}

Abukawa \etal\cite{Abukawa-ecoss22-03} noted that RHEED images taken as a function of azimuthal angle and assembled, as previously described, into intensity maps as a function of the scattering vector provide access to a large volume of reciprocal space.  These scattering vectors span the full reciprocal space above the sample, except for a small funnel shaped region around the axis of rotation (see \fref{abukawa}).  This large volume of reciprocal space provides intensity data over the CTRs and any SR present, and as such enables the possibility of using intensity data to extract atomic positions from the RHEED data.   Abukawa \etal\cite{Abukawa-ecoss22-03} presented an initial data set from azimuthal scan RHEED data of a Si(111)-$\sqrt{3}\times\sqrt{3}$-Ag surface, and used the Patterson function to see changes in atomic positions due to a surface structural phase transition at 130K.  

Romanyuk \etal\cite{Romanyuk-cjp-06} used the Patterson method to study the formation of iron-silicide from two mono-layers of Fe deposited on Si(111) from azimuthal scan RHEED data.  To interpret the data they calculated the Patterson function for a set of reference structure models of the Fe-Si surface and compared them with the experimental data. They were able to discount several models for the surface atomic structure, but they found spots in their experimental Patterson function, which might have been artifacts,  that they could not reproduce with the models.  

In a second paper,  Abukawa \etal\cite{Abukawa-prl-06}  noted the analogy of the RHEED data collected as a function of azimuthal angle with that of the Weissenberg camera method for XRD.  They also pointed out that using standard kinematic-based x-ray diffraction tools, such as the Patterson method, for structural determination with RHEED data which clearly includes inelastically scattered structure is problematic.   In conventional dynamical analysis of RHEED intensities,  inelastic scattering issues are minimized by carefully choosing the incident beam conditions and which diffraction spots to study.  With azimuthal scan RHEED data, this is not possible, so instead  Abukawa \etal\cite{Abukawa-prl-06} used a high pass filter set to 5.5 eV below the electron beam energy to remove the most significant effects of multiple scattering.  This enabled a surface crystallographic analysis on a Si(111)-$\sqrt{3}\times\sqrt{3}$-Ag surface, via the Patterson function, to produce atomic positions in excellent agreement with the known atomic structure.  

The interatomic distances obtained directly via the Patterson function, a kinematic diffraction tool, from electron diffraction images provides a very valuable new tool to explore the surface atomic structure of epitaxial grown oxides and the surface upon which they are grown.  Furthermore, the removal of inelastically scattered electrons from the RHEED images, by energy filtering, allows quick access to more interpretable data over a large area of reciprocal space.   In addition, the energy selection of the scattered electrons also allows access to basic chemistry information at the surface via Auger events.

\section{\insitu\ Kikuchi diffraction}

Removing the inelastic scattering effects from the densely collected RHEED images in order to apply kinematic based diffraction methods does, however, remove a significant part of the remaining crystallographic data that can be generated by diffraction in the RHEED geometry.  In addition to the Bragg reflections of the RHEED beam, typical RHEED images of high quality samples also show Kikuchi lines generated by inelastic scattering processes.  For many decades Kikuchi patterns have been used by electron microscopists to precisely orient samples in 3-dimensional space as they contain a representation of the full 3-dimensional crystal structure.  Much of the work surrounding their use can be directly applied to  \insitu\ and real-time pattern collected while doing RHEED.  This section provides some background on the technique and shows its potential to help distinguish subtle crystal structure changes that could be occurring due to the epitaxial growth of complex oxides.

\begin{figure}
\begin{center}
\includegraphics[width=0.87\textwidth]{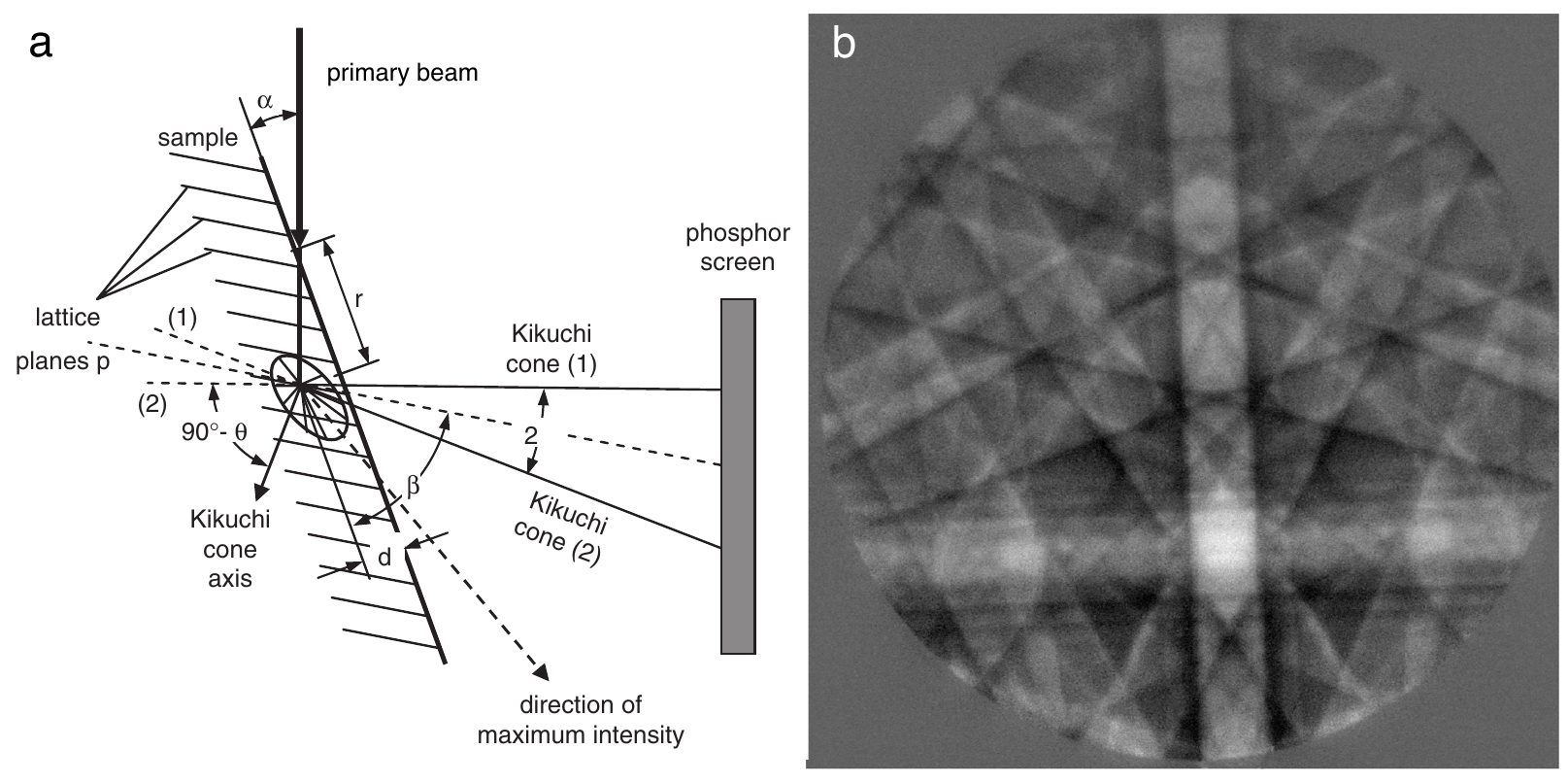}
\caption{ a) Scheme of backscatter Kikuchi pattern formation: the primary beam enters the sample under the angle $\alpha$. The primary electrons are incoherently scattered with an intensity distribution indicated by the ellipsis, and subsequently coherently scattered by the lattice planes $p$ into pairs of Kikuchi cones (1) and (2) with an opening angle of $180^o - 2\theta$ ($\theta$ is the Bragg angle) around the cone axis. b) A typical Kikuchi pattern of Si taken with 15 KeV electrons in a field-emission gun Scanning Electron Microscope.  Reprinted from \cite{Zaefferer-u-07} \label{Zaefferer}}

\end{center}
\end{figure}

In the 1930's, while studying thin mica films using an electron beam, Kikuchi\cite{kikuchi-1} observed a background structure, as well as the expected diffraction peaks (for a historical review, see for example, Ref. \cite{dingley-4}).  This background structure consisted of a series of parallel line pairs, later called Kikuchi lines, which were interpreted in a ``two-event model''.  The first step of the model involved the incoming collimated and mono-energetic beam of electrons being diffused in the crystal by an unspecified incoherent scattering process.  In essence, this first step generates a point source of electrons traveling in multiple directions, inside the material. The second step then involves standard Bragg diffraction of the diffused electrons from the planes in the material.  From any point inside the crystal there are two possible angles for which Bragg diffraction from a single set of lattice planes can occur.  Therefore, each set of crystallographic planes generate two diffraction cones with a large opening angle of $180^o - 2\theta$ and an axis perpendicular to the diffracting lattice planes.   These cones will be seen on a collecting screen as a pair of lines,  {\em i.e.} the Kikuchi lines (see \fref{Zaefferer}a).  Zone axes are defined as the intersection line between a set of crystallographic planes, and since the Kikuchi lines are a projection of the crystallographic planes, any point where the various Kikuchi line pairs intersect will be a zone axis.  

In reality, the multiple scattering process required to generate the Kikuchi pattern leads to more complex contrasts in the image than simple pairs of lines.\cite{reimer-book}  The diffraction pattern is better described as bands with excess or deficient Kikuchi lines that define the edge of the band (see \fref{Zaefferer}b).  Deal \etal \cite{Deal-u-08}  studied the energy dependence of the Kikuchi patterns and observed that contributions to the Kikuchi patterns arise from electrons that have up to about 80\% of their incident energy.  The Kikuchi band contrast is found to be maximum for electrons with 97\% of their incoming energy, or having lost 450eV of energy with respect to the original 15 KeV beam.

These Kikuchi patterns (also called Electron Backscatter Diffraction (EBSD) patterns)  are used extensively in Scanning Electron Microscopes (SEM) and Tunneling Electron Microscopes for sample alignment and texture maps based on crystallographic information (see, for example, Ref \cite{basic-SEM/TEM-kikuchi-reference}).  During RHEED studies, Kikuchi lines are often used to demonstrate qualitatively that the long range order of the sample is good.  They have also been used quantitatively to determine the average crystal potential and the crystal misorientation.\cite{braun-book}    However, these Kikuchi diffraction patterns contain a representation of the full three-dimensional structure of the probed material, and can therefore be used to determine the space group and potentially the 3-dimensional atomic positions of an epitaxially grown film.  To make that practical, the Kikuchi patterns need to be collected away from the main Bragg diffraction.

Baba-Kishi\cite{babakishi-rheed} first demonstrated the combined collection of RHEED and Kikuchi patterns, using two different phosophor screens, within an SEM to study a thin GaAs capping layer deposited on GaAlAs.    Horio \cite{horio-jssn-06} also investigated the combination of RHEED and Kikuchi patterns on a Si(111)-$7\times 7 $ surface.  In this case Horio used a conventional RHEED setup and added a second hemispherical screen in line with the sample normal.  This hemispherical screen, which lead to the name astrodome RHEED, was used to enable an acceptance angle of $95^o$ and to minimize the distortion of the Kikuchi patterns that occur due to the standard gnomonic projection of the patterns when they are collected on a flat screen.  Procedures for the the collection and analysis of Kikuchi patterns from spherical surfaces have recently been discussed.\cite{Day-jm-08}

\begin{figure}
\begin{center}
\includegraphics[width=0.57\textwidth]{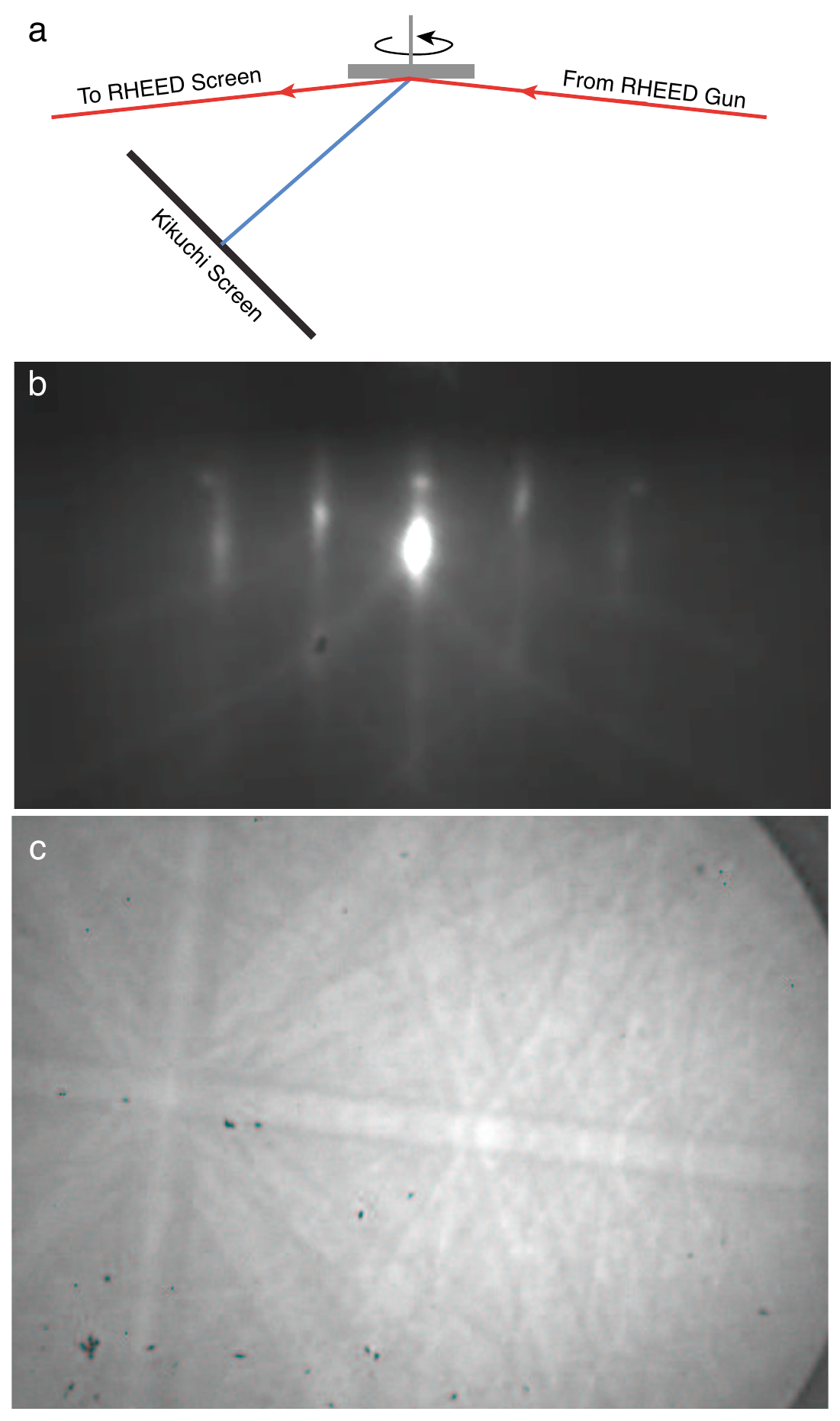}
\caption{ (a) The two phosphor screen geometry to allow simultaneous RHEED and Kikuchi pattern collection. The simultaneous (b) RHEED pattern and (c) Kikuchi pattern of a
\STO\ substrate at 650\oC\ and under a $1 \times 10^{15}$
atoms/cm$^2$ sec flux of atomic oxygen.
\label{sto-substrate-650}}
\end{center}
\end{figure}

The particular location of Horio's hemispherical screen does not allow for the use of this screen during film growth.  In \fref{sto-substrate-650}a we show a slightly altered geometry, with a flat screen, that allows the Kikuchi screen to be monitored throughout the film growth processes to provide \insitu\ and real-time data collection.  The electron beam has an angle of incidence on the sample of about 2\o, as is standard for RHEED.  The Kikuchi screen is 5 cm away from the sample with an angle of 45\o\ to the sample normal.  \Fref{sto-substrate-650}b and c show simultaneous RHEED and Kikuchi images collected from a (001) \STO\ substrate after heating to 650\oC\ under atomic oxygen.  The strongest Kikuchi band contrast is found with the incident angle of the incoming electron beam at around 15\o\ with this current setup.  Zaefferer \cite{Zaefferer-u-07} notes that maximum electron intensity is obtained when the angle of incidence of the electron beam and the angle of emission of the Kikuchi patters are similar.  This suggests that placing a screen to capture the Kikuchi patterns with an angle to the sample normal of as close to 90\o\ as possible, while still avoiding the main   Bragg diffraction peaks, would be advantageous.  However, as discussed later, the best location for the screen will depend on the particular sets of Kikuchi bands that provide access to the most significant structural information.

One major concern with using the Kikuchi patterns for the structural analysis of thin films is the influence of diffraction from the substrate on the observed pattern.  If the film is too thin, the electron beam will interact with the substrate as well as with the film.  In general, 30 keV electrons, as used in this work, have a penetration depth of roughly 500\ \AA.  This is then decreased to about 50\ \AA\ because of the low angle of incidence of the electron beam with the sample.  In agreement with this, C.J. Harland \etal\cite{harland} determined an information depth of $\leq 100$\ \AA, while Kohl\cite{kohl} calculated an information depth of about 50-60\ \AA\ for the reciprocally-related electron channeling patterns.  This approximate probe depth of 50--100\ \AA\ is further experimentally supported by work of Baba-Kishi and Dingley\cite{baba-kishi-2} on bulk NiS$_2$.    These results imply that a film thickness of $>200$\ \AA\ will be adequate to avoid the influence of the substrate on the Kikuchi pattern formation.

To gain qualitative information from the Kikuchi patterns requires two important pieces of geometrical information: the location of the pattern center ({\em i.e.} the electron-source point), and the specimen-to-screen distance.  Three methods for this are outlined in the literature: conic fitting\cite{biggin,venables-2}, the circular mask technique\cite{venables-2}, and the known orientation method\cite{baba-kishi-4}.  Due to the fact that epitaxial growth occurs on a well characterized single crystal substrate, the known orientation method is the easiest and most accurate method. 

Once the geometrical information is known, the analysis of the Kikuchi pattern to obtain space group information\cite{baba-kishi-1,baba-kishi-2,dingley-4}  starts with the analysis of individual zone axes ({\em i.e.} points where at least two sets of Kikuchi lines cross), to define their respective point group symmetries.  The combination of these point group symmetries from several zone axes will then allow the determination of the crystallographic point group.\cite{dingley-4}  From this point on, the determination of the space group relies on the ability to calculate some rough lattice parameters from interzonal angles from which to simulate the Kikuchi diffraction pattern for comparison with the experimental pattern.  These procedures have been fully automated in most of the EBSD units attached to SEMs.

\insitu\ Kikuchi diffraction of epitaxial thin films allows several simplifications over the general procedure outlined above for analyzing the images.  The first is that there is normally prior knowledge of the expected structure, so that a full {\em ab-inito}  structure determination is not needed. The second is that the diffraction pattern of the substrate, when collected immediately prior to growth, can be directly compared to the as-grown film to determine changes in structure and to define the exact geometry.  Finally, the azimuthal rotation of the sample to collect RHEED data, also allows the collection of the large portion of the Kikuchi pattern that is generated in the hemisphere above the sample, thus allowing many more zone axes to be seen.   In our experimental geometry we can see $2/3$ of the full hemisphere, although due to the flat nature of the phosphorus screen the edges of our images are geometrically distorted.

\begin{figure}
\begin{center}
\includegraphics[width=0.57\textwidth]{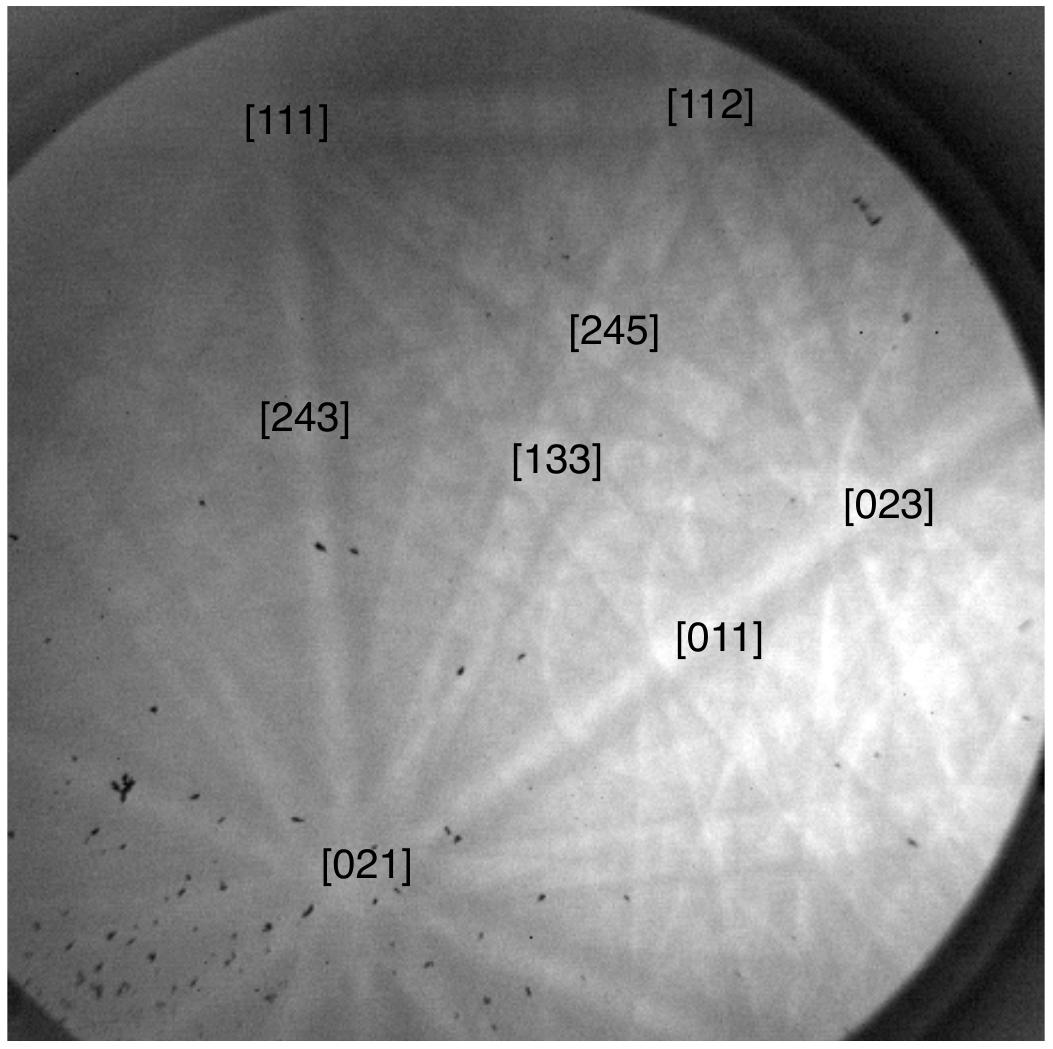} 
\caption{ The Kikuchi pattern of a SrRuO$_3$ film grown on a \STO\ substrate at 650\oC\ and under a $1 \times 10^{15}$
atoms/cm$^2$ sec flux of atomic oxygen; zone axes are indicated.
\label{sro-indexed-image}}
\end{center}
\end{figure}

\Fref{sro-indexed-image} shows a Kikuchi pattern collected after the growth of a SrRuO$_3$ film on SrTiO$_3$ (see Ref. \cite{Klein-jpcm-96} and \cite{ingle-apl-99} for growth details) while the film is still at 650\ \oC\ and under a $1 \times 10^{15}$ atoms/cm$^2$ sec flux of atomic oxygen.  Several zone axes have been indexed according to the tetragonal unit cell expected at 650\ $^{\circ}$C.\cite{Vailionis-apl-08}  The image quality of these particular Kikuchi patterns is of similar quality to many of the original patterns collected in SEMs.  There has been significant improvements in hardware for the collection of these patterns in SEM instruments, and all of those improvements could be applied to the \insitu\ collection of images from thin film growth chambers.  In fact, it is theoretically possible to directly install an EBSD unit, built for an SEM, directly into a thin film growth chamber.

The basic unit cell can be obtained by locating various zone axes on the images, and determining their relative geometry, after accounting for any geometrical image distortion.   Lattice plane spacing, or d-spacing, can then be determined by measuring the angular separation of the Kikuchi line pairs, and using the Bragg equation.  However, as pointed out by Wilkinson\cite{Wilkinson-u-96}, for 20 keV electrons the Bragg angle for low order reflections is quite small and the edge of the Kikuchi band is not sharp, which allows only a few parts in 100 accuracy.   The improvement in this accuracy for higher order Kikuchi lines is hindered by their lack of clarity in the image.

In SEM based Kikuchi pattern measurements, Wilkinson\cite{Wilkinson-u-96} demonstrated that it is possible to obtain sensitivity to strains of the order of 0.02\% on thick Si$_{1-x}$Ge$_x$ epitaxial layers grown on planar Si substrates.  This was done by careful measurements and analysis of interzonal angles.  This method is limited to strains that distort the shape of the unit cell, although that is of primary interest in understanding the tetragonal distortion that occurs with epitaxial growth.  However, changes in atomic positions may not be readily detected by these analysis techniques.  Therefore we are also interested in looking at the differences in the intensity of specific Kikuchi bands as a route to explore other epitaxially influenced changes in the structure of thin films.

\begin{figure}
\begin{center}
\includegraphics[width=0.57\textwidth]{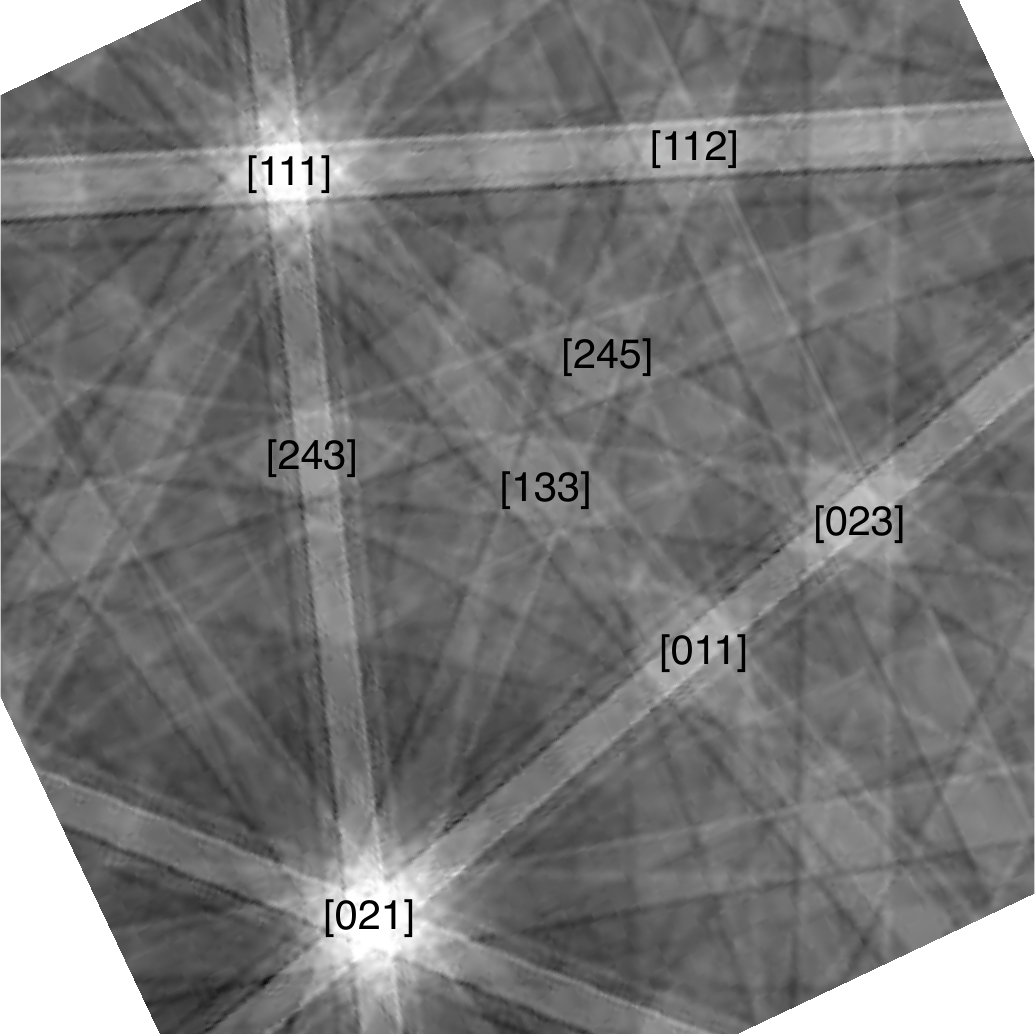}
\caption{ The calculated Kikuchi pattern for tetragonal SrRuO$_3$ with the same orientation as the experimental image of \fref{sro-indexed-image}.
\label{SrRuO3-tetragonal-simulation}}
\end{center}
\end{figure}

In order to sort out what to look for in the experimental Kikuchi pattern to help distinguish possible changes in the crystal structure of a film due to effects of epitaxy, we implemented a simulation based on a pixel by pixel application of the kinematic model of diffraction.  A kinematic simulation will generate the correct geometry of the images, but will not generate correct intensities as Kikuchi pattern formation is a multi-scattering phenomena, and therefore requires a full dynamical diffraction model to predict intensities.  However, Zaefferer\cite{Zaefferer-u-07} argues that the kinematical  intensities are not too far off the experimental ones, and that with the use of a simple intensity correction procedure based on a two-beam diffraction theory, the simple simulation can generate qualitatively good results.  We have implemented these corrections, as outlined in Zaefferer's work, to simulate Kikuchi patterns from various structural models of interest.  \Fref{SrRuO3-tetragonal-simulation} shows a simulation centered on the [133] zone of tetragonal SrRuO$_3$, calculated for indices up to 15, which can be compared directly with the experimental image in \fref{sro-indexed-image}.  

An example of a possible epitaxial influence on the structure of a grown film is the potential buckling of the Cu-O planes in the high-pressure phase of SrCuO$_2$ when grown with compressive strain.    SrCuO$_2$ is the end member ($n=\infty$) of the homologous high-pressure series Sr$_{n-1}$Cu$_{n+1}$O$_{2n}$,\cite{Hiroi-jssc-91}.  Due to the simplicity of the crystal structure and the planar nature of its Cu and O ordering, it is often called the parent compound of the high temperature superconductors\cite{Jorgensen-prb-93} and is considered to be a 2-dimensional quantum magnet.\cite{Yasuda-prl-05}     An extremely large change in the $c$-axis of this material was seen when it was grown by epitaxy with reduced $a$ and $b$ lattice parameters.\cite{Ingle-jap-02}   One possible explanation for this anomalous expansion in the $c$-axis is the buckling of the Cu-O planes, suggested by Vailionis \etal.\cite{Vailionis-prb-97}   The proposed buckling of the Cu-O planes occurs by the motion of just the oxygen atoms, and as such it is extremely hard to see in x-ray diffraction.  However, Kikuchi diffraction is an ideal tool to look at the potential of an epitaxially driven change in the dimensionality of the Cu-O planes.

\begin{figure}
\begin{center}
\includegraphics[width=0.77\textwidth]{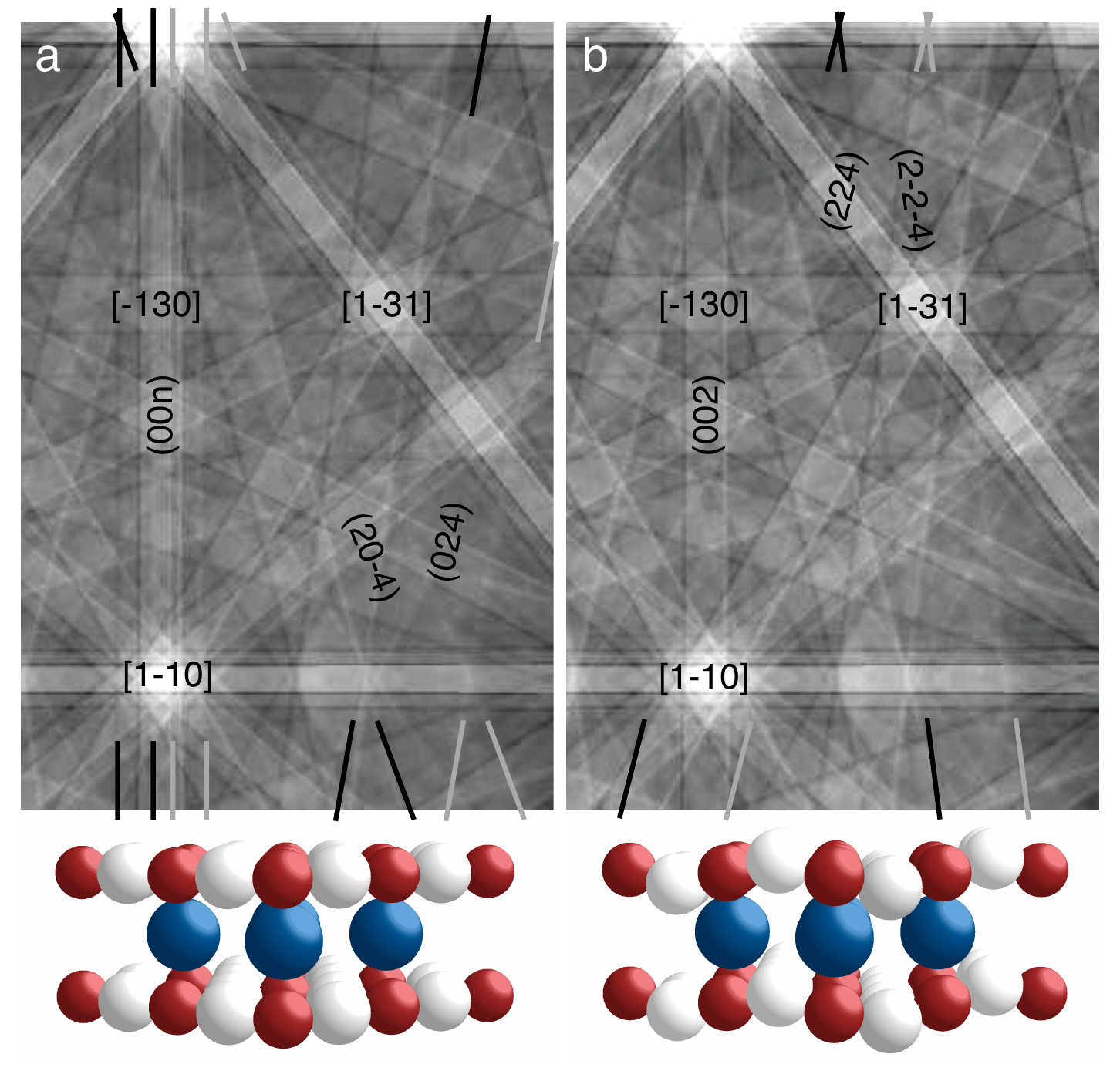}
\caption{ The calculated Kikuchi patterns, and structures, of the high pressure phase of SrCuO$_2$ without buckled Cu-O planes (a), and with buckled Cu-O planes (b).  The light and dark lines at the top and bottom of the image are guides to the eye of the respective excess and deficient lines that define the edges of the Kikuchi bands that change between the two images. 
\label{SrCuO2-comparison}}
\end{center}
\end{figure}

\Fref{SrCuO2-comparison} shows several changes that are expected in the Kikuchi pattern, centered around the ($\overline{1}30$) zone axis, if the Cu-O planes were to buckle by the oxygen atoms moving in and out of the $a$-$b$ plane by 10\% of the $c$ lattice constant (\Fref{SrCuO2-comparison}b) in contrast to a planar Cu-O structure (\Fref{SrCuO2-comparison}a).  In this part of the Kikuchi pattern there is an increase in intensity of the scattering from the \{224\} planes -- indexed according to the enlarged unit cell required to describe buckling -- due to buckling (highlighted in \fref{SrCuO2-comparison}b).  Furthermore, there is also significant reduction in intensity expected for the (001) and (003) planes, as well as the cubic-indexed \{204\} planes for the buckled structure (highlighted in \fref{SrCuO2-comparison}a).

There are a number of other expected changes in the Kikuchi pattern for the buckling of the Cu-O plane in SrCuO$_2$, but primarily in the high-index planes.  Although high-index planes are present in the image, they not only have inherently a lower intensity but are overlaid with the more intense low-index planes, and locations in the pattern are not always present that allow them to be seen.    However, these simulations suggest that even subtle changes in the structure of epitaxially grown films are quite distinguishable in RHEED generated Kikuchi patterns.  In addition, the ability to collect a Kikuchi pattern from the well characterized substrate just prior to the growth of a film provides a remarkable opportunity to minimize unknown experimental and geometrical effects in the patterns and clarify the significant aspects of the simulations.  This affords a tremendous advantage to the final analysis and interpretation of the Kikuchi patterns collected from the  epitaxial film.

Recent work on multi-beam dynamic models, by Wilkelmann \etal\cite{Winkelmann-u-08}, which include improvements in the handling of inelastic scattering processes and the description of the excitation process show very good agreement between simulated and experimental patterns of GaN\{0001\}.   The simulations have reproduced the fine structure around zone-axes and higher-order Laue zone rings seen in the experiments.    These results, along with continual improvements in hardware\cite{day-jom-08} and software that are being applied to EBSD in SEMs, allow a much more accurate and fine-grained approach to inspecting and understanding the experimental Kikuchi patterns collected from thin epitaxial film to look for subtle changes in the atomic structure.

\section{Conclusion}

The use of high energy electrons at grazing incidence during the growth of epitaxial films generates scattered electrons which can be analyzed to obtain an abundance of information about the morphology and the crystallographic structure of the film.   The collection of the standard RHEED patterns at all azimuthal angles of rotation, and ideally with energy filtering to isolate just the elastic scattering, can be reconstructed in real-time to provide immediate visualization of the symmetry of the crystal truncation rods, highlighting any surface reconstructions that may be occurring, as well as the basic in-plane lattice parameters.  The intensity from this type of data set can also be treated with the Patterson function approach of x-ray diffraction to help reveal the atomic positions of atoms on the surface.  Furthermore, the electrons that undergo inelastic scattering to generate Kikuchi patterns, also allow access to the space group and atomic positions of the thin epitaxial film.  These Kikuchi patterns can also be collected in real-time, and simultaneously with the RHEED data on a separate screen.  These various methods to collect and analyze the scattered electrons from RHEED allow a remarkably complete set of data which help elucidate the full crystallographic structure of the potentially novel structures, interfaces, and surfaces that can be formed by the influence of epitaxy on the growth of thin films.

\ack{This work was supported by NSERC and a DAAD scholarship (MP).  NI thanks M.R. Beasley for the use of the Molecular Beam Synthesis chamber to explore the collection of the Kikuchi patterns, and thanks I.S. Elfimov for valuable discussion during this work.}

\section*{References}

\end{document}